\documentclass[preprint,3p,11pt]{elsarticle}

\usepackage{graphicx,subfigure}
\usepackage{amssymb}
\usepackage{amsmath,mathrsfs}
\usepackage{bm}
\usepackage{url}
\usepackage{xcolor}
\usepackage{lipsum}
\usepackage{comment}

\journal{Journal of Computational Physics} 

\numberwithin{equation}{section}

\usepackage{orcidlink}

\usepackage{tabularx}

\usepackage{algpseudocode}
\PassOptionsToPackage{hyphens}{url}\usepackage{hyperref} 

\begin{document}

\begin{frontmatter}

\title{Neural network sampling of Bethe-Heitler process in particle-in-cell codes}

\author[GOLPIST]{Óscar Amaro \orcidlink{0000-0003-0615-0686} \corref{cor}}
\ead{oscar.amaro@tecnico.ulisboa.pt}
\cortext[cor]{Corresponding author}
\author[GOLPIST]{Chiara Badiali \orcidlink{0000-0001-6450-7511}}
\author[GOLPIST]{Bertrand Martinez \orcidlink{0000-0003-4344-7657}}

\address[GOLPIST]{GoLP/Instituto de Plasma e Fus\~ao Nuclear, Instituto Superior T\'ecnico, Universidade de Lisboa, Lisbon, Portugal}

\begin{abstract}
This study uses neural networks to improve Monte Carlo (MC) implementations of the Bethe-Heitler process in Particle-In-Cell (PIC) codes.
We provide a neural network that is as accurate as pre-calculated tables, and requires a hundred times less memory to store.
It is trained to predict Bethe-Heitler pair production cross-sections for atomic numbers 1-50 and photon energies between 1 MeV and 10 GeV in the PIC code \texttt{OSIRIS}.
We first validate our approach against a theoretical estimate in a simplified context.
We later prove that both approaches have similar performance in a typical relativistic laser-plasma interaction scenario.
The large memory decrease accessible with neural networks will enable introducing more advanced cross-section models for Bethe-Heitler pair production and other QED mechanisms in the MC modules of PIC codes.
\end{abstract}


\end{frontmatter}

\section{Introduction}

A growing number of countries are building facilities hosting high-power lasers \cite{JPCSHernandez2010,HPLSEPapadopoulos2016,OEYoon2019,HPLSEBromage2019,HPLSERadier2022}. 
The extreme states of matter driven by PetaWatt-class systems are prone to trigger Quantum Electrodynamics (QED) processes mediated by the Coulomb field of ions such as the emission of high-energy photons via Bremsstrahlung \cite{RMPKoch1959} and electron-positron ($e^-/e^+$) pair generation via the Bethe-Heitler mechanism \cite{RMPMotz1969}. 
The experimental investigation of these two effects finds applications in the development of bright gamma-ray sources for radiography of dense objects \cite{RSIPerry1999,PoPCourtois2013} and in the creation of relativistic $e^-/e^+$ pair plasmas \cite{PRLChen2009,PRLSarri2013,PoPChen2023} that are essential to grasp the dynamics of magnetic fields and the acceleration of particles in an astrophysical context.
Such ambitious experiments at the forefront of science require advance preparation with numerical simulations.
Modern Particle-In-Cell (PIC) codes \cite{BookBirdsall1991} are ideal to addressing this challenge as they can model relativistic laser-plasma interactions, as well as QED effects through a Monte Carlo (MC) module \cite{ASSHenderson2011,PoPYan2012,EPJDJiang2014}.
Improving the MC module of PIC codes thus marks a major milestone toward bringing better and more accurate predictions in the context of forthcoming experiments at high-power laser facilities.

Monte Carlo modules within PIC codes have been actively developed for several decades and are now a standard numerical tool to model collisions and binary processes \cite{takizukaBinaryCollisionModel1977, nanbuTheoryCumulativeSmallangle1997, nanbuWeightedParticlesCoulomb1998, sentokuNumericalMethodsParticle2008, peanoStatisticalKineticTreatment2009, perezImprovedModelingRelativistic2012,PREAlves2021}.
Although the algorithm could collide every particle with each other, it would be too computationally expensive.
Instead, the algorithm can either arrange particles by pairs when they are close in phase-space \cite{takizukaBinaryCollisionModel1977}, or determine an emissivity or an opacity from the local plasma conditions \cite{PoPSentoku2008,PRERoyle2017}.
In both cases, it evaluates the cross-section of the processes and uses the result to mediate energy exchanges, particle production and annihilation.
The total and differential cross-sections of a mechanism are stored in pre-calculated tables to avoid a direct and costly evaluation at run-time \cite{grismayerLaserAbsorptionQuantum2016}.
Besides look-up tables, there are relatively few alternatives explored in the literature.
Chebyshev polynomial approximations work well for data with homogeneous features, and have recently been used to fit particle production rates in MC codes \cite{nielsenGPUAcceleratedMonte2022, montefioriSFQEDtoolkitHighperformanceLibrary2023}.
This approach is demonstrated to be accurate, and provides a continuous value for the cross-section whereas a table does not and requires an additional interpolation step.

However, cross-section models implemented so far in PIC codes are simplified as compared to the theory available in the literature.
In particular, Coulomb processes are described as a function of the energy of the incident and emitted particles in most MC modules \cite{PoPSentoku2008, moritakaPlasmaParticleincellSimulations2013, nakamuraNumericalModelingQuantum2015, martinezHighenergyRadiationPair2019,LPBNakamura2015,HPLSEWu2018,JPCSMoritaka2013,PPCFVyskocil2018}.
This approach does not yet account for the role of the angle of the incoming and created particles that are described theoretically \cite{RMPKoch1959,RMPMotz1969}.
In addition, the need to describe the screening of the atomic potential in a hot and dense plasma for Coulomb processes requires introducing additional parameters to the cross-section as the plasma temperature, density and ionization degree \cite{martinezHighenergyRadiationPair2019}.
The high-amplitude electromagnetic fields that plasmas can sustain are also expected to influence the cross-section of atomic processes \cite{PLBKrachkov2019}.
We can however note that tables rapidly become impractical to fit such cross-sections as their number of dimensions would be of the order of five to ten.
Even with polynomials, the number of coefficients needed for 3D or higher-dimension functions would still require extensive look-up tables.
The current limits observed with tables and polynomials should be overcome in order to provide more suitable models in numerical simulations of relativistic laser-plasma interactions.
It is therefore crucial to suggest new approaches to evaluate cross-sections within PIC codes.


Neural networks are gradually being introduced into plasma simulation codes.
They have been recognized for their efficiency, accuracy and compact models for regression tasks \cite{cholletDeepLearningPython2018, geronHandsOnMachineLearning2019}.
For instance, this capability is exploited to develop surrogate models for laser-driven ion acceleration~\cite{djordjevicModelingLaserdrivenIon2021,schmitzModelingLiquidLeaf2023a}, for hybrid particle accelerator beamlines~\cite{sandbergSynthesizingParticleinCellSimulations2024a}  or to represent the static local field correction in path integral MC simulations in the whole warm dense matter regime \cite{dornheimStaticLocalField2019}.
Neural networks can replace a specific part of a simulation code.
As an example, neural networks can be used to advance particles in hybrid codes~\cite{wuLearningEfficientHybrid2022}.
In Ref.~\cite{kluthDeepLearningNLTE2020}, neural networks are implemented in the hydrodynamic-radiative code \texttt{HYDRA} to infer spectral opacities in non-local thermal equilibrium.
The speed up and accuracy in the evaluation of spectral opacities is crucial in the context of Inertial Confinement Fusion.
Regarding PIC codes, the \texttt{OSIRIS} framework implements neural networks evaluating Compton cross-sections \cite{badialiMachinelearningbasedModelsParticleincell2022}, displaying a comparable precision and a better run-time as the direct calculation.

We show in this work how to exploit neural networks to enhance the MC implementation of the Bethe-Heitler process in PIC codes.
The neural network we present stands out by its low memory storage, which is a hundred times smaller than the pre-calculated table for the same accuracy.
Leveraging this large memory decrease for cross-section models available with neural networks, our work paves the way to introduce more sophisticated cross section models for Bethe-Heitler and other QED processes in PIC codes. 

To explain our approach, we first detail what parts of the typical Monte Carlo module are replaced by neural networks in section \ref{sc:method}.
In particular, the first neural network evaluates the total cross-section of the Bethe-Heitler process, while the second samples the energy of the emitted positron.
Section \ref{sc:datasets} explains the method to create the dataset and the training of the neural network.
From this training, we identify the best performing neural networks.
For comparable accuracy, the neural network that samples the positron energy has only 327 trainable parameters, while the table requires 40000 values, implying a difference in memory storage of two orders of magnitude.
In section \ref{sc:results} we first validate both implementations (with the table and with the neural network) by comparing successfully the positron spectra with a theoretical estimate.
In a second step, we performed a realistic simulation of relativistic laser-plasma interactions where an intense laser pulse drives pair production during its interaction with a thin foil.
The results indicate that the spectra of positrons obtained with the neural network approach is as accurate and efficient as with the acknowledged table method.
Finally, we present our conclusions and the prospects for MC modules in PIC codes in section \ref{sc:conclusions}.

\section{Method}\label{sc:method}

In this section we describe the machine learning approach followed in this project, detailing the numerical frameworks used, the structure of the neural networks, and the algorithms implemented in \texttt{OSIRIS} to compute the particle production probabilities.

\begin{figure}[thpb]
    \centering
    \includegraphics[width=380pt]{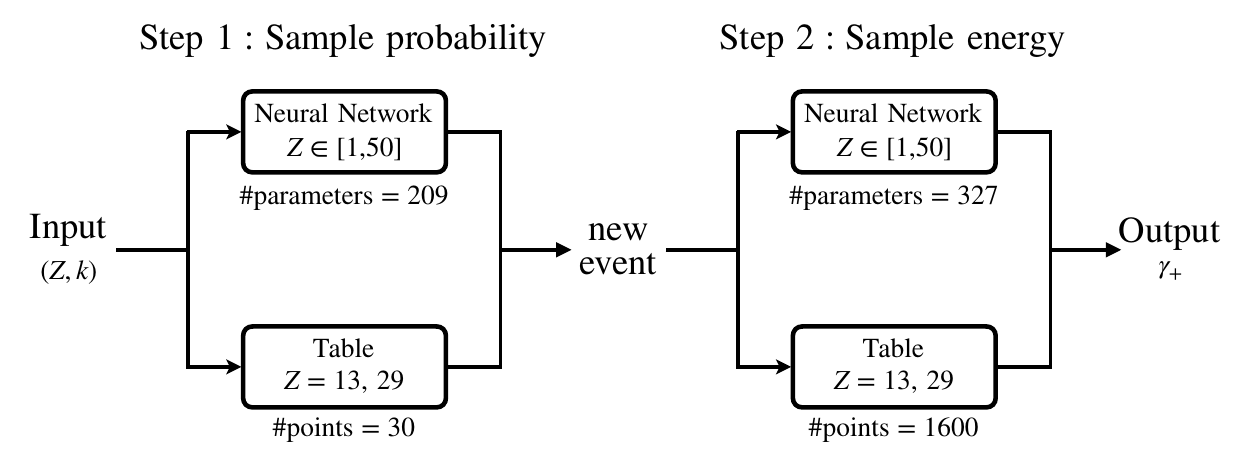}
    \caption{Graph with algorithms implemented in \texttt{OSIRIS} for computing probability of Bethe-Heitler electron-positron pair production and the energy sampling of the created particle. To describe $Z\in [1,50]$ atoms, the table approach would require 750 points for the total cross-section and 40000 points for the differential cross-section, whereas the Neural Networks trained in this work only require 209 and 327 parameters, respectively, for comparable accuracy. Therefore, the amount of memory needed to store the neural network is thus a hundred times smaller than the table.}
    \label{fig:graph_algorithm}
\end{figure}

Contrary to \cite{badialiMachinelearningbasedModelsParticleincell2022}, we are not replacing the Monte Carlo (MC) routine with a conceptually different approach. Instead, we substitute the routine computing the total cross-section ($\sigma_{\mathrm{BH}}$) and the inverse cumulative distribution function (CDF$^{-1}$) with tables by neural networks.
Our approach is sketched in Fig.~\ref{fig:graph_algorithm}.

We use dense neural networks, where each node of one layer is connected to all nodes of the previous layer. 
We train networks with varying number of inputs, and hidden layers (less than 4), but always with a 1D output, since the goal is to learn scalar probability distribution functions. 
We use the \texttt{Keras} framework \cite{chollet2015keras} to build and train the neural network. Furthermore, we use the \texttt{Python} libraries \texttt{Numpy}, \texttt{Pandas} and \texttt{SciPy}, and the \texttt{C++} \texttt{boost} scientific library \cite{BoostLibraries} to generate and visualize the datasets. Training is carried out until the loss (a measure of the distance between the prediction of the network and the theoretical values) has decreased significantly and stabilized.
To link the Python-trained neural network models to the fortran physics code \texttt{OSIRIS}, we used the packages \texttt{neural-fortran} and \texttt{FKB} \cite{
curcicModernFortranBuilding2020, ottFortranKerasDeepLearning2020}.

In this work, the Bethe-Heitler cross-section model assumes the atom is neutral and isolated. This means the total cross-section depends on two parameters: the atomic number $Z$ and the energy of the incident photon $k=\hbar\omega/mc^2$. The differential cross-section, and therefore the cumulative distribution function, depends on these two parameters plus the Lorentz factor of the positron generated denoted $\gamma_+$. By definition the CDF is monotonic; however this behavior is not explicitly enforced in the neural network. If the accuracy of the model is sufficiently high, this behavior can be accounted for.

For both cross sections, the activation function between layers is Rectified Linear Unit function (ReLU), the output node has a sigmoid activation function and the optimizer used was Adam (for example, see \cite{chollet2015keras}). In the case of the total cross-section, the loss function and the evaluation metric are Mean Absolute Percentage Error (MAPE) and binary cross entropy, respectively, while for the CDF$^{-1}$ these are MAPE and mean absolute error (MAE). Usual machine learning methods were found to improve the training of the neural network, namely data balancing, batch normalisation and increasing the number of entries of the dataset (see section \ref{sc:datasets}).

Additionally, it was observed that the neural network of the CDF$^{-1}$ could not accurately reproduce the data when $\gamma_+$ and the CDF were close to $0$.
Since the CDF$^{-1}$ is an odd-function around $\gamma_+=k/2$, the information contained in the intervals $\gamma_+/k \in [0,0.5]$ and $\gamma_+/k \in [0.5,1]$ is the same. Therefore, we split the datasets into these two cases, and verified that the trained models performed better in the interval $[0.5,1]$. Increasing the complexity of the neural network would probably mitigate this issue further, albeit at the expense of slower run time.

\section{Data generation and model training}\label{sc:datasets}

In this section we explain how datasets for the training of the model were generated and report the training of the networks. Furthermore, we describe how the datasets were prepared and balanced, two standard techniques in machine learning \cite{cholletDeepLearningPython2018}.

\subsection{Training Dataset}

Contrary to \cite{badialiMachinelearningbasedModelsParticleincell2022}, our training data does not originate from \texttt{OSIRIS} simulations. Instead, simple scripts compute the total and differential cross sections. The scripts generating the data are written in \texttt{Python} and \texttt{C++}.
They are used to produce datasets of approximately 1 million entries for the two cross-sections.
The first columns represent the input parameters, namely the atomic number $Z$, the incident photon energy $k=\hbar \omega/mc^2$ and the positron Lorentz factor $\gamma_+$.
The last column is the target value we would like to predict, \emph{i.e.} the total or differential cross-section. To attempt a more representative dataset over the input domain, variables were generated from the grids: $Z \in [1, 50]$ (linearly spaced), $k \in [2, 2\times 10^4]$ (logarithmically spaced) and $\gamma_+/k \in [0.5,1]$ (linearly spaced).
We use the symmetry of the Bethe-Heitler cross-section to reduce the interval of the normalized Lorentz factor $\gamma_+/k$ from $[0,1]$ to $[0.5,1]$.
It is worth noting that most of the PIC simulations conducted with Coulomb processes in the past decades fall within this extremely large range of atomic number and photon energies.

Over this input domain the total cross-section can vary between $10^{-32}$ and $10^{-26}~\mathrm{m}^2$ in real units. This broad dynamic range is detrimental to train the NN, for this reason we instead normalize the data to span the interval $[0,1]$.
More specifically, we take the negative logarithm of the cross-section, subtracting the minimum and then dividing by the range. In contrast, the CDF (and consequently the CDF$^{-1}$) is by definition already in the interval $[0,1]$, and the quantity $\gamma_+/k$ as well.  We note that for all parameters the curve CDF$^{-1}$ is always close to the function $f(\gamma_+/k)=\gamma_+/k$, which makes it difficult for the model to distinguish between different input combinations. Instead, we create a new variable $\Delta = \mathrm{CDF}^{-1}(\gamma_+/k)-\gamma_+/k$, such that the model learns this difference. We proceed to normalize this variable to the range $[0,1]$ as detailed in \ref{app:prep}.

After this step, we proceed to balance the statistics of the datasets. Some target values can occur orders of magnitude more often than other values, which leads the neural network to disregard the statistically rare entries and to perform poorly on the entire input domain. We use a simple algorithm to reduce the frequency of the over-represented entries, thereby flattening the histogram of the target in the dataset.

\subsection{Training of neural network models}

The cross validation is performed by randomly splitting the data into training (70\%), validation (15\%), and testing (15\%) groups with a fixed random seed.

\begin{figure}[thpb]
    \centering
    \includegraphics[width=420pt]{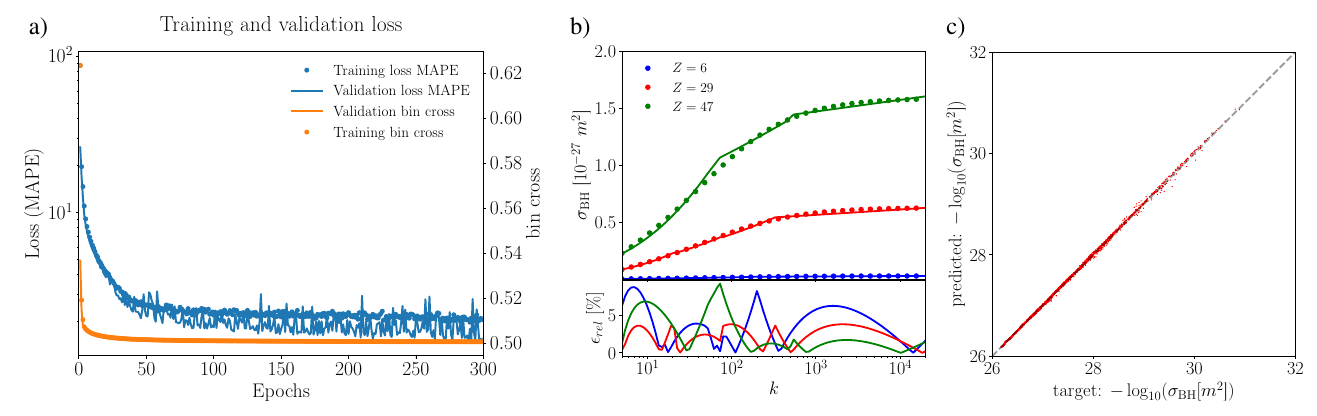}
    \caption{Results for the total cross section: a) Training iterations with loss function MAPE and metric binary cross-entropy, b) Evaluating the neural network model prediction against theory for different photon energies $k$ and atomic number $Z$ (dots - theory, line - neural network), c) Evaluating the neural network model prediction against theory for subset of training data.}
    \label{fig:ModelTCS}
\end{figure}

In the case of the total cross-section $\sigma_{\mathrm{BH}}$, the layout with two layers, 12-12 nodes presents the best results. 
On the left panel of Figure \ref{fig:ModelTCS}, we show the evolution of the training and validation loss throughout training. After 300 epochs, the losses reduce significantly and show convergence of training. On the middle panel, we show for three different materials / values of $Z$ and as a function of photon energy $k$ the comparison between the trained total cross-section (full line) against the theoretical expression used in to produce the training data (in dots). On the lower subplot the absolute relative error for each $Z$ value. On the right panel we show a scatter plot of the neural network's prediction against the theoretical / target value of the training data. A 1:1 correspondence and small spread around this line indicate that the neural network is both accurate and precise. 
On the whole interval $Z\in[1,50]$ and $k\in[2,2\times 10^4]$, we obtain an average relative error on the order of $5\%$ for the total cross section $\sigma_{\mathrm{BH}}$.

\begin{figure}[thpb]
    \centering
    \includegraphics[width=420pt]{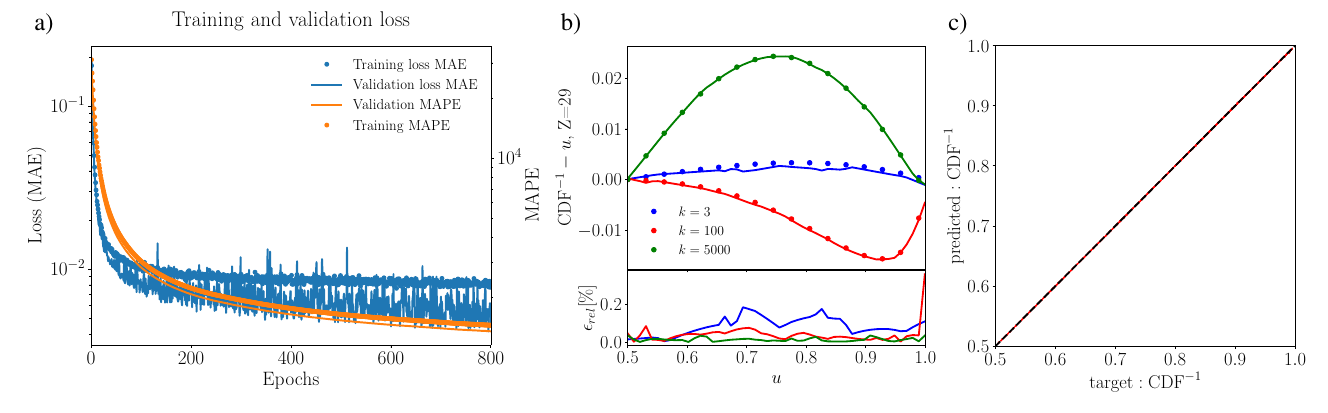}
    \caption{Results for the differential cross section: a) Training iterations with loss function MAE and metric MAPE, b) Evaluating the neural network model prediction against theory for different positron to photon energies  $k/\gamma_+$ and different photon energies $k$ (dots - theory, line - neural network), c) Evaluating the neural network model prediction against theory for a subset of training data.}
    \label{fig:ModelCDF}
\end{figure}

In the case of the inverse of the cumulative distribution function CDF$^{-1}$, the best performing architecture has 3 layers, 8-16-8 nodes. Results of the training of the network are shown in Fig. \ref{fig:ModelCDF}.
Similarly to the total cross-section case, the left panel indicates convergence of the training.
The middle panel shows good agreement between theory and the inferred values by the model for three $Z$ values.
The right panel proves there is an almost 1:1 correspondence between target and model prediction.
On the whole interval $Z\in[1,50]$, $k\in[2,2\times 10^4]$ and $\gamma_+/k\in[0.5,1]$ we obtain an average relative error below $0.2\%$.

The neural network for the total cross-section $\sigma_{\rm BH}$ has 209 trainable parameters to describe all atomic numbers in the range $Z\in[1,50]$ while the CDF$^{-1}$ has 327. This contrasts with the look-up table approach, which requires 800 parameters per atomic number for comparable accuracy, amounting to a total of 40000 parameters for the entire range $Z\in[1,50]$. Thus, replacing the tables with neural networks led to reduce the memory required to store the cross-section model by two orders of magnitude.

\section{Results}\label{sc:results}

In this section we present results from \texttt{OSIRIS} simulations with both the table and the neural network approaches.

\subsection{Validation of the table and neural network implementations against theory}

In the first case, we benchmark the Bethe-Heitler cross-sections in an idealized setup, consisting of a monoenergetic population of photons colliding against a copper target ($Z = 29$). The normalizing simulation frequency is $\omega_p = 1.88\times 10^{15}~\mathrm{s}^{-1}$, corresponding to a normalizing density of $n_p=1.11 \times 10^{21} ~\mathrm{cm}^{-3}$; photons have an energy of $k=2000$, there are $20096$ macro-photons per cell with density of $10^{-4} ~n_p$ extending on a length $L_k=8 ~c/\omega_p$ the ions have a density of $100~n_p$; , the time step is $\mathrm{dt} = 0.002~\omega_p^{-1}$, and simulation only lasts one time step to avoid multiple scattering events and for the energy of the pairs created not to change over time.
All other Monte-Carlo / radiative processes are turned off to ensure a clear comparison.

In Figure \ref{fig:osirisNN_TCSiCDF} a) we compare the theoretically expected final positron energy spectrum against the look-up-table and neural network approaches.
The theoretical number of positrons in one time step is provided by the following formula
\begin{equation} \label{eq:BHdiff}
    \mathrm{d}N_+/\mathrm{d}\gamma_+ = n_k L_k \times \mathrm{d}\sigma_{\mathrm{BH}}/\mathrm{d}\gamma_+ n_i c ~\Delta t
\end{equation}
where $n_k L_k$ is the number of gamma-rays per unit of surface, and $\sigma_{\mathrm{BH}} n_i c$ the frequency of positron emission for one gamma-ray. The differential cross-section $\mathrm{d}\sigma_{\mathrm{BH}}/\mathrm{d}\gamma_+$ is defined in Eq.(25) from Ref \cite{martinezHighenergyRadiationPair2019}.
From this, it can be computed that the relative error in positron yield of the look-up table approach is $1.06 \%$ and for the neural network it is $3.71 \%$. We observe this larger error is due to the slight under-estimation of the total cross-section by the first neural network.
Despite this, we note the other neural network sampling the positron energy reproduces very well the shape of the theoretical spectrum.

\subsection{Production run 2D}

In a second setup, we test the long-time accuracy of the neural network approach by running a more typical 2D simulation for laser-target interaction setup with Coulomb-QED processes.
In this setup, an intense laser pulse interacts with an aluminium target ($Z=13$) to produce and accelerate positrons, which are not initially present in the simulation.
It propagates along the x-axis and is linearly polarized along the y-axis.
The normalizing simulation frequency and density have the same values as in the previous subsection.
The aluminium target is initially fully ionized with a thermal plasma temperature of 1 keV.
It has a thickness of $6.28 ~c/\omega_p$ for the uniform density part (both ions and electrons have a density of $20~n_p$), and an exponentially varying pre-plasma of the same scale-length.
The incident laser has a vector potential of $a_0=200$ and frequency $\omega_L=\omega_p$, it has a transverse Gaussian focusing with a spot size of $W_0=20.11 ~c/\omega_p$, and a Gaussian temporal envelope with Full-Width at Half Maximum (FWHM) $\tau=62.8 ~\omega_p^{-1}$.
The simulation domain has dimensions $288\times 384 ~(c/\omega_p)^2$ in longitudinal and transverse directions respectively. The time step is $0.030 ~\omega_p^{-1}$ and the simulation is run for $300 ~\omega_p^{-1}$. To save computational resources, we use a moving window in the longitudinal direction x.
We allow leptons to lose energy through nonlinear Inverse Compton scattering in the quantum regime, which here is the primary mechanism for photon emission.

\begin{figure}[thpb]
    \centering
    \includegraphics[width=350pt]{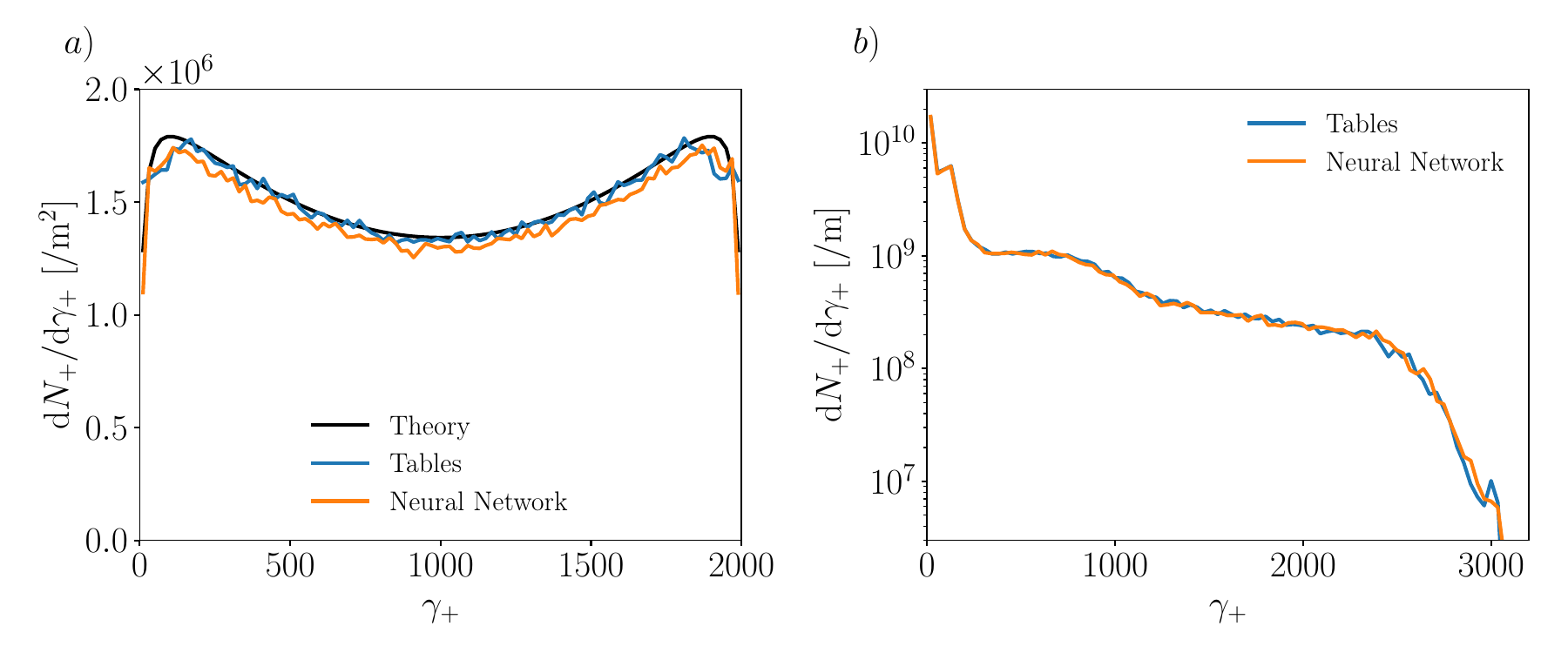}
    \caption{Positron spectrum for various simulations. a) Positron spectra from a monoenergetic ($1 \, \rm GeV$) photon beam incident on a copper target ($Z=29$), simulated for single time step in 1D setup. The black curve represents the analytical spectrum (see equation \ref{eq:BHdiff}), the blue curve is the spectrum using pre-computed tables, and the orange curve corresponds to using neural networks for the total cross-section ($\sigma_{\rm BH}$) and the inverse of the Cumulative Distribution Function (CDF$^{-1}$). b) Final positron spectrum for an intense laser pulse ($I \simeq 5\times 10^{22} \, \rm Wcm^{-2}$) propagating through a pre-expanded thin foil ($Z=13$) in the Relativistic Self-Induced Transparency regime (RSIT) \cite{kawRelativisticNonlinearPropagation1970}. The approach with neural network has the same accuracy as tables.}
    \label{fig:osirisNN_TCSiCDF}
\end{figure}

During the interaction, the intense laser pulse accelerates some electrons to relativistic velocities. As they oscillate in the high-amplitude field, electrons radiate gamma-rays via the nonlinear Inverse Compton scattering process. The gamma-rays emitted forward travel through the high-density plasma and can produce electron-positron pairs via the Bethe-Heitler mechanism. We choose a regime where the plasma is heated and expands on a time-scale shorter than the laser duration, known as the Relativistic Self-Induced Transparency regime \cite{kawRelativisticNonlinearPropagation1970}. As a result, the laser propagates through the expanding plasma, heats it further and sustains the longitudinal static field accelerating positrons to energies reaching 1.5 GeV.

In Figure \ref{fig:osirisNN_TCSiCDF} b) a good agreement is shown between the table and neural network approaches. Despite the statistical fluctuations on the higher energy part of the spectrum, we verify that the average absolute relative error between the two approaches was $1.4\%$, thus confirming that the neural network approach can accurately be deployed in production runs.

\begin{table}[ht]
\caption{Run-time for a production run where the Bethe-Heitler cross-sections are sampled by different methods.}
\label{tab:table1}
\centering
\begin{tabular}{cccc}
\hline
 & \textbf{Simulation 1} & \textbf{Simulation 2} & \textbf{Simulation 3} \\
 & (Only tables) & (Hybrid table / network) & (Only neural networks) \\
\hline
Total time $\rm (s)$ & $2997$ & $2981$ & $5526$ \\
\hline
\end{tabular}
\end{table}

We now compare the performance of our QED-PIC simulations relying on the table and neural network approaches.
To measure the efficiency, we choose to report the total run-time of the simulations from Fig.~\ref{fig:osirisNN_TCSiCDF} in Tab.~\ref{tab:table1}.
In simulation 1, lasting $2997 \, \rm s$, the total cross-section $\sigma_{\rm BH}$ and the inverse Cumulative Distribution Function $ \rm CDF^{-1}$ are both read in tables.
In simulation 2, running for $2981 \, \rm s$, we introduce a hybrid approach where $\sigma_{\rm BH}$ is still read from a table but $\rm CDF^{-1}$ is now computed by a neural network.
The two timings are very similar because the routine evaluating $ \rm CDF^{-1}$ is rarely called during the simulation.
Indeed, the Bethe-Heitler pair production probability per time step is of the order of $10^{-4}$ or lower for typical relativistic laser-plasma interaction setups.
The relative difference we measure between the performances of simulation 1 and 2 is only of $0.5\%$ and is not significant.
It is reasonable to expect a small difference in performance since the final positron spectrum of the two systems differ by a small amount ($1.4\%$).
However we note that simulation 3 has a significantly longer run-time of $5526 \, \rm s$.
This growth can be explained as the total cross-section $\sigma_{\rm BH}$ is now computed with a neural network.
Indeed, the evaluation by the neural network requires multiplying matrices whereas a table interpolation involves basic operations on reals.
This difference at the routine-level becomes significant for the total simulation time because the total cross-section $\sigma_{\rm BH}$ is evaluated at all time steps and for all particles.

\section{Conclusion}

\label{sc:conclusions}

Our work demonstrates the potential of neural networks as a powerful tool for enhancing the MC implementation of the Bethe-Heitler process in PIC codes.
The neural network we identify maintains the same level of accuracy as a pre-calculated table, yet it takes up a hundred times less storage in memory.
In our approach, we provide a detailed description of all the steps we followed to obtain this outcome.
We explain how the training data is generated and how the model is trained.
The best neural network identified reproduces with accuracy the theoretical positron spectra expected in a simple scenario where mono-energetic photons propagate in a cold target.
Plus, we perform a typical production run of pair production in relativistic laser-plasma interactions and observe the neural network is as accurate and efficient as the table to predict the final positron spectra.

The main prospect of this work is to 
leverage the drastic drop in memory storage offered by neural networks as compared to the established approach with tables.
This advantage will enable implementing more refined cross-section models for the Bethe-Heitler pair production and other QED processes in PIC codes.
For instance, one can think of effects in a strong electromagnetic field, fusion reactions, collisions as well as atomic physics and radiative transport.
However, we expect neural networks to be best suited when tables and polynomials become too large in memory, namely when the cross-section or the rate is a function of three of more parameters. 
The second prospect is to enhance the performance of neural network evaluation in the Monte Carlo module.
One promising approach is to port the computations performed by the neural network from CPU to GPU, resorting to OpenACC directives or utilizing Cuda Fortran.

\section*{Data availability statement}

The code used for this project is openly available in a git repository: \url{https://github.com/OsAmaro/osirisBetheHeitlerML}. The training data and set of the final model weights, are available in a zenodo repository: \url{http://doig.org/10.5281/zenodo.11422851}

\section*{Conflict of interest}

The authors declare no conflict of interest.

\section*{Acknowledgments}

This article comprises part of the PhD thesis work of Óscar Amaro, which will be submitted to Instituto Superior Técnico, University of Lisbon.

The authors would like to acknowledge fruitful discussions with GoLP team members and for proofreading the manuscript, in particular Prof. Marija Vranic, Dr Lucas I. Inigo Gamiz, Mr. Diogo Carvalho, Mr. Pablo Bilbao, and Mr. Lucas Ansia. 

\'O. Amaro acknowledges the support of the Portuguese Science Foundation (FCT) Grant No. UI/BD/153735/2022.
Chiara Badiali acknowledges the support of the Portuguese Science Foundation (FCT) Grants No. PRT/BD/152270/2021 DOI 10.54499/PRT/BD/152270/2021.
B. Martinez acknowledges the support of the Portuguese Science Foundation (FCT) Grants No. CEECIND/01906/2018 and PTDC/FIS-PLA/3800/2021 DOI: 10.54499/PTDC/FIS-PLA/3800/2021.

We acknowledge the EuroHPC Joint Undertaking for awarding this project access to the EuroHPC supercomputer LUMI, hosted by CSC (Finland) and the LUMI consortium through a EuroHPC Regular Access call.
Some simulations were performed at the IST cluster (Lisbon, Portugal).

\begin{appendix}

\section{Pre- and post-processing of neural network output}\label{app:prep}

In this section we summarize the steps taken to prepare the data before training and to recover the values after training. After training of the model, the output of the neural network needs to be translated back to real units, reversing the data preparation step.

\noindent
The training dataset for the total cross-section has 9 million entries: 3000 point along the $Z$ axis and 3000 points along the $k$ axis.
The cross-section is normalized between 0 and 1 in order to enhance the efficiency of the training.
We therefore define the target $\hat{\sigma}$:
\begin{equation}
\hat{\sigma} = \frac{-\log_{10}(\sigma)+\log_{10}(\sigma_{\mathrm{min}})}{-\log_{10}(\sigma_{\mathrm{max}})+\log_{10}(\sigma_{\mathrm{min}})}
\end{equation}
where $\sigma_{\mathrm{min}}=10^{-26.173498}$ and $\sigma_{\mathrm{max}}=10^{-32.151504}$.

\noindent
The training dataset for the inverse of the cumulative distribution function has ~10 million entries: 216 points along the $Z$ axis, 216 points along the $k$ axis and 216 points along the $\gamma_+/k$ axis.
Due to the symmetry of the Bethe-Heitler cross-section, the cumulative distribution function (and its inverse) are close to the curve $f(\gamma_+/k)=\gamma_+/k$.
For this reason, we define a target that is not directly $\mathrm{CDF}^{-1}(\gamma_+/k)$ but rather the difference $\Delta=\mathrm{CDF}^{-1}(\gamma_+/k)-\gamma_+/k$ and then we normalize it between 0 and 1.
In other words, the target is 
\begin{equation}
\hat{\Delta} = \frac{\Delta - \Delta_{min}}{\Delta_{max}-\Delta_{min}}
\end{equation}
where $\Delta_{min}=-0.018344$ and $\Delta_{max}= 0.026995$.

\end{appendix}

\bibliographystyle{elsarticle-num}
\bibliography{bibtex}

\end{document}